\documentstyle[twocolumn,aps,psfig]{revtex}

\begin{document}
\draft
\twocolumn[\hsize\textwidth\columnwidth\hsize\csname @twocolumnfalse\endcsname
\title{Ferromagnetic, A-type, and Charge-Ordered CE-type States \\
in Doped Manganites using Jahn-Teller Phonons}

\author{Seiji Yunoki, Takashi Hotta, and Elbio Dagotto}

\address{National High Magnetic Field Lab and Department of Physics,
Florida State University, Tallahassee, FL 32306}

\date{\today}
\maketitle

\begin{abstract}
The two-orbital model for manganites with both non-cooperative and 
cooperative Jahn-Teller phonons is studied at hole density x=0.5 
using Monte Carlo techniques. 
The phase diagram is obtained varying the electron-phonon coupling 
and the $\rm t_{2g}$-spins exchange. 
The insulating CE-type charge- and orbital-ordered state 
with the $z$-axis charge $stacking$ observed in narrow-bandwidth manganites
is stabilized in the simulations. 
Its charge gap $\Delta_{\rm CO}$ is much larger than the critical 
temperature $\rm k_B T_{CO}$.
Metallic-like A-type and ferromagnetic states are also obtained in the same 
framework, and the phase boundaries among them have first-order 
characteristics.
\end{abstract}


\vskip2pc]
\narrowtext

The complicated interplay between charge, spin, orbital, and 
lattice degrees of freedom is believed to induce the unexpected magnetic and 
transport phenomena observed in Mn-oxides, such as 
the `colossal' magnetoresistance (MR) effect\cite{jin}.
The subtle properties of manganites are especially complex at 50\% hole 
density where experiments have established that an insulating 
charge-ordered (CO) state exists in 
$\rm Nd_{0.5} Sr_{0.5} Mn O_3$\cite{PSMO},
$\rm Pr_{0.5} Ca_{0.5} Mn O_3$\cite{tomioka},
$\rm La Sr_2 Mn_2 O_7$\cite{li},and several others.
The state has alternating $\rm d_{3x^2-r^2}$/$\rm d_{3y^2-r^2}$
orbital and CE-type antiferromagnetic (AF) order\cite{goodenough}.
Its melting by a magnetic field leads to a huge negative MR\cite{tokura}, 
probably induced by mixed-phase tendencies\cite{science,previous}.

In spite of its importance, the origin of the x=0.5 CO-state is still unclear.
The observation of charge populating only the even (or odd) Mn-ion 
sublattice in the xy plane naively suggests that the nearest-neighbor (NN) 
Coulomb repulsion V$_{\rm NN}$ is important for its stabilization.
However, the $z$-axis stacking of charge\cite{wollan} and 
existence of bistripes at x$>$0.5\cite{bistripes}, both penalized
by a strong NN-repulsion, show that V$_{\rm NN}$ is smaller than expected
and other ingredients are needed to understand the CO-state\cite{moritomo}.
In addition, at x=0.5 experiments exhibited two other phases in competition:
(i) A $\rm d_{x^2-y^2}$ orbital-ordered (OO) but charge-disordered (CD) state 
with A-type AF state and two-dimensional (2D) metallicity
(in, e.g., $\rm Pr_{0.5} Sr_{0.5} Mn O_3$\cite{PSMO}).
(ii) A metallic spin ferromagnetic (FM) CD and orbital-disordered (OD) state
(in, e.g., $\rm La_{0.5} Sr_{0.5} Mn O_3$).
For a dominant V$_{\rm NN}$, stable CD states at x=0.5 are difficult to 
understand. 
It may occur that electrons coupled only through on-site Coulomb interactions 
produce this exotic behavior, as suggested by Hartree-Fock 
calculations \cite{mizokawa}.
However, in this context unbiased many-body calculations are difficult and
computational studies have technical complications such as the sign-problem.
In addition, the `colossal' oxygen isotope shift of the CO transition in 
$\rm Nd_{0.5} Sr_{0.5} Mn O_3$ (NSMO)\cite{zhao} cannot be explained 
using purely Coulombic approaches, which predict no isotope effect.

Searching for an alternative mechanism to stabilize the charge-stacked 
CO-state of manganites, in this paper the first comprehensive computational 
analysis of the half-doped two-orbital model with both non-cooperative and
cooperative Jahn-Teller (JT) phonons in the large
electron-phonon coupling ($\lambda$) regime is presented. 
Our main result is that a CE-type CO orbital-ordered ground state can be 
stabilized in this framework with charge properly stacked along the $z$-axis.
Many other features found in the simulations are also in excellent 
correspondence with experimental observations. 

The two-orbital Hamiltonian is given by
\begin{eqnarray}
&& \rm H \! = \! - \! \sum_{{\bf i} {\bf v} \sigma a b} t^{\bf v}_{ab}
c^\dagger_{{\bf i} a \sigma} c_{{\bf i+v} b \sigma}
\! -J_H  \! \sum_{\bf i} {\bf S}_{\bf i} \cdot {\bf s}_{\bf i} 
\! + \! J_{AF} \sum_{\bf \langle ij \rangle} 
{\bf S}_{\bf i}\cdot {\bf S}_{\bf j}  \nonumber \\ 
&& \rm 
\!+ \ \!\lambda \sum_{{\bf i} a b \sigma} c^{\dagger}_{{\bf i} a \sigma} 
(Q_{2{\bf i}}\sigma_1+Q_{3{\bf i}}\sigma_3)_{\rm ab}
c_{{\bf i} b \sigma} \!
\! + {1 \over 2} \! \sum_{\bf i} ( {Q}^2_{2{\bf i}} + {Q}^2_{3{\bf i}}) \!,
\end{eqnarray}
where $\rm c_{{\bf i}1 \sigma}$ ($\rm c_{{\bf i} 2 \sigma}$) is the
destruction operator for an e$_{\rm g}$-electron with spin $\sigma$
in the $\rm d_{x^2-y^2}$ ($\rm d_{3z^2-r^2}$) orbital at site ${\bf i}$,
${\bf v}$ is the vector connecting NN sites, 
and $\rm t^{\bf v}_{ab}$ is given by 
$\rm t_{\rm 11}^{\bf x}$=$-\sqrt{3}{\rm t}_{\rm 12}^{\bf x}$
=$-\sqrt{3}{\rm t}_{\rm 21}^{\bf x}$=$\rm 3{\rm t}_{\rm 22}^{\bf x}$=t
(energy unit) for $\rm {\bf v}$=${\bf x}$,
$\rm t_{\rm 11}^{\bf y}$=$\sqrt{3}{\rm t}_{\rm 12}^{\bf y}$
=$\sqrt{3}{\rm t}_{\rm 21}^{\bf y}$=$\rm 3{\rm t}_{\rm 22}^{\bf y}$=t
for $\rm {\bf v}$=${\bf y}$, 
and $\rm t_{\rm 22}^{\bf z}$=4t/3,
$\rm t_{\rm 11}^{\bf z}$=$\rm t_{\rm 12}^{\bf z}$=$\rm t_{\rm 21}^{\bf z}$=0 
for $\rm {\bf v}$=${\bf z}$.
The Hund coupling $\! \rm J_H \!$ ($>$0) links the $\rm e_g$-electron spin 
$\! \rm {\bf s}_{\bf i}\! $=$\! \rm \sum_{a \alpha \beta} \!
\rm c^\dagger_{{\bf i}a \alpha} \! {\bbox \sigma}_{\alpha \beta} \!
c_{{\bf i}a \beta}\!$
and the localized $\rm t_{2g}$-spin $\rm {\bf S}_{\bf i}$ assumed classical 
($\!\rm |{\bf S}_{\bf i}|\!$=1).
The $\rm e_g$-electron density $\! \rm \langle n \rangle \!$ 
varies with a chemical potential $\mu$
(the hole density is x=1$\!-\! \rm \langle n \rangle \!$).
$\rm J_{AF}$ is the AF coupling between NN $\rm t_{2g}$-spins.
The fourth term couples $\rm e_g$-electrons to JT-mode distortions, 
$\rm Q_{2{\bf i}}$ and $\rm Q_{3{\bf i}}$, 
assumed classical \cite{previous,millis}.
$\sigma_1$ and $\sigma_3$ are Pauli matrices.
When the  $\rm \{ Q \}$ are independent, ``non-cooperative'' phonons
are used \cite{previous,millis}.
In real materials, however, adjacent $\rm MnO_6$ octahedra share an oxygen. 
In this ``cooperative'' case, the proper variables are 
the oxygen displacements $\rm u_{\bf i}^{\bf v}$ from equilibrium 
along the Mn-Mn bond in the $\rm {\bf v}$-direction, and extra terms are 
added to Eq.~(1) for the breathing-mode phonon\cite{hotta2}.
Standard Monte Carlo (MC) simulations for the classical spins and phonons 
are here used\cite{previous}.

Figure 1(a) already illustrates one of the main results of the present study.
Shown are the MC real-space charge correlations at x=0.5 for
representative couplings in 2D at low temperature(T). 
For intermediate $\rm J_{AF}$ ($0.1 \alt \rm J_{AF} \alt 0.2$),
the charge correlations were found to be robust at all distances and they are
positive (negative) on the even (odd) sites, 
compatible with the CO-pattern observed in x=0.5 experiments.
Our results show that this behavior is essentially independent of $\lambda$ 
for intermediate $\rm J_{AF}$, although the CO-pattern is more 
easily observed for large $\lambda$.
For small $\rm J_{AF}$ ($\alt 0.1$), on the other hand, 
the CO-state appears only for large $\lambda$.
The observed CO-state originates from the tendency of the JT-distorted 
$\rm Mn^{3+}$-ions to maximize their relative distances 
to exploit the kinetic energy of the mobile carriers.
Figure 1(b) contains the Fourier transform of the $\rm t_{2g}$ spin 
correlations $\rm S({\bf q})$ in the CO-regime.
For intermediate $\rm J_{AF}$, peaks at ${\bf q}=(\pi,0)$ and 
$(\pi/2,\pi/2)$ are observed in excellent agreement with the result of 
a perfect CE-spin arrangement.
Note that size effects appear small in these calculations, and 
throughout the paper.
For small $\rm J_{AF}$, only one peak is detected at ${\bf q}=(0,0)$,
indicating the FM phase. 
Under the CE-type spin environment, it can be shown that
the CO-state is stable even for small $\lambda$, since 
the system is effectively one-dimensional (1D) due to 
charge confinement along the zigzag FM-chain \cite{hotta}.
In this sense, $\rm J_{AF}$ plays a role more important than $\lambda$ 
for the CE-state stabilization.
This is in contrast to the FM state where a finite and 
large $\lambda$ is needed 
to stabilize the CO-state due to its higher dimensionality.

\begin{figure}[htbp]
\vspace{-1.0cm}
\centerline{\psfig{figure=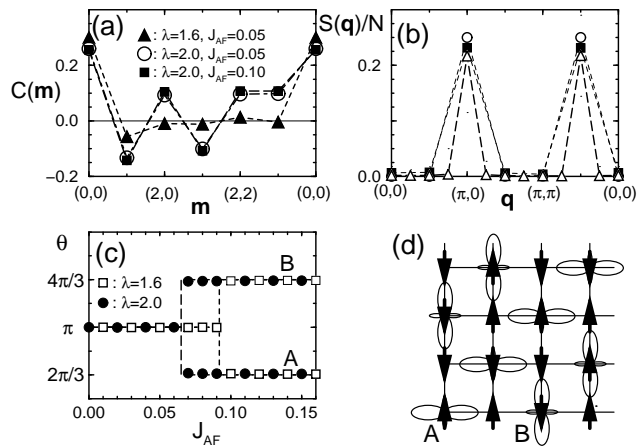,width=9.50cm,angle=-90}}
\vspace{0.2cm}
\caption{
(a) Charge correlation $\rm C({\bf m})$=$\rm 
\langle n_{\bf i} n_{\bf i+m} \rangle$-$\rm \langle n \rangle^2$ 
with $\rm n_{\bf i}$=$\rm \sum_{a\sigma}
c^\dagger_{{\bf i}a\sigma} c_{{\bf i}a\sigma}$
for $\rm J_H$=$\infty$, T=1/100, and x=0.5, using 
a 4$\times$4 cluster with non-cooperative phonons and
periodic boundary conditions.
The squares (circles) [triangles] 
were obtained at $\rm J_{AF}$=0.1, $\lambda$=2.0 (0.05, 2.0) [0.05, 1.6]. 
(b) Spin structure-factor
$\rm S({\bf q})$=$\rm \sum_{\bf m} e^{i {{\bf q}\cdot{\bf m}} }
\langle {{{\bf S}_{\bf i} }\cdot{ {\bf S}_{\bf i+m}}} \rangle$
for the $\rm t_{2g}$-spins at the same parameters of the squares in (a). 
The squares (circles) here denote the 4$\times$4 MC (ideal CE-type state) 
results. 
Triangles are MC results on an 8$\times$8 cluster.
(c) $\rm \theta_{\bf A}$ and $\rm \theta_{\bf B}$ (see text) 
vs. $\rm J_{AF}$ for $\lambda$=1.6 and 2.0.
(d) Charge, spin and orbital arrangement compatible with our MC results.
}
\vspace{-0.1cm}
\end{figure}

Regarding the orbital arrangement, let us consider sites with substantial 
charge (A and B in Fig.~1(d))\cite{comm6}.
The occupied orbital at each site is given by 
$\rm |{\tilde 2}\rangle_{\bf i}\!=\!-\sin (\theta_{\bf i}/2)\!|1\rangle
\!+\! \cos (\theta_{\bf i}/2)\!|2\rangle$ 
with $\rm \theta_{\bf i}\!=\!\tan^{-1}\!
(\langle Q_{2{\bf i}} \rangle \! / \!\langle Q_{3{\bf i}} \rangle)$.
In Fig.~1(c), $\theta_{\rm A}$ and $\theta_{\rm B}$ vs. $\rm J_{AF}$ 
are plotted for $\lambda$=1.6 and 2.0.
For small $\rm J_{AF}$, 
$\theta_{\rm A}\!=\!\theta_{\rm B}\!=\!\pi(\rm d_{x^2-y^2}\!)$, 
since the orbitals prefer to have large overlaps both in the 
x- and y-directions.
On the other hand, for intermediate $\rm J_{AF}$, 
$\theta_{\rm A}\!=\!2\pi/3(\rm d_{3x^2-r^2}\!)$ and 
$\theta_{\rm B}\!=\!4\pi/3(\rm d_{3y^2-r^2}\!)$.
To understand this result, note that the orbital is polarized 
along the hopping direction to improve the kinetic energy. 
Due to this effect, the zigzag hopping path in the CE-type CO-state induces 
the alternating $\rm \!d_{3x^2-r^2}\!/\! d_{3y^2-r^2}\!$ orbitals.
The above results in charge, spin, and orbital correlations 
for intermediate $\rm J_{AF}$ are schematically shown in Fig.~1(d),
which is the CE-type CO-state observed in experiments.

\begin{figure}[htbp]
\vspace{-0.9cm} 
\centerline{\psfig{figure=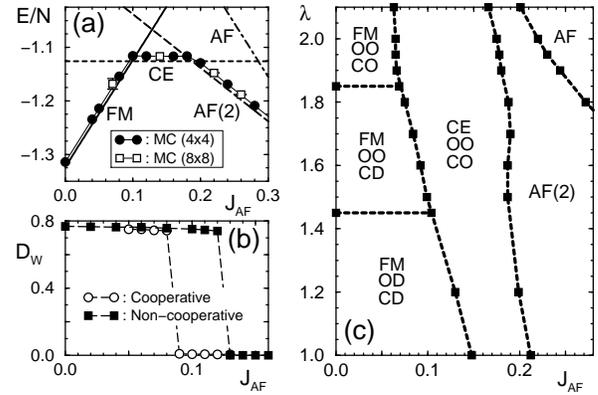,width=8.5cm,angle=-90}}
\vspace{0.2cm}
\caption{
(a) MC energy per site E/N vs. $\rm J_{AF}$ at x=0.5, $\lambda$=1.5,
T=1/100, and $\rm J_H$=$\infty$ with non-cooperative phonons. 
The FM state is charge-disordered and orbital(uniform)-ordered. 
``AF(2)'' is explained in Ref.~\protect\cite{kink}.
``AF'' is AF in both directions.
The straight lines correspond to the 
finite $\rm J_{AF}$ correction to $\rm J_{AF}$=0 MC results, when a
perfect spin configuration is assumed.
(b) Drude weight $\rm D_W$ vs. $\rm J_{AF}$ at T=1/100, x=0.5, 
using a $4\times4$ cluster. 
The results with cooperative (non-cooperative) JT-phonons were
obtained at $\lambda$=1.5 (1.2). 
(c) Phase diagram of the 2D two-orbital model with non-cooperative
phonons at low-T.
The phase boundaries were obtained as in (a) 
using 4$\times$4 and 8$\times$8 clusters. 
All transitions are of first-order. 
In the FM-phases, the OO-state is uniform in the $\rm d_{x^2-y^2}$
orbitals, while in the CE-phase the pattern Fig.~1(d) was observed.
}
\vspace{-0.1cm}
\end{figure}

Figure 2(a) shows a typical MC energy vs. $\rm J_{AF}$ for $\lambda=1.5$.
At small $\rm J_{AF}$, a FM-state is stabilized 
since even at large $\lambda$ the optimal background for electrons 
has all spins aligned. 
However, its energy is penalized by the AF exchange and at realistic values 
of $\rm J_{AF}$ an equal AF-FM mixture such as the CO CE-state minimizes 
the energy.
The discontinuity in $\rm d(E/N)/dJ_{AF}$ signals a {\it first-order} 
transition, as in experiments.
With increasing $\rm J_{AF}$, the CE-state is 
destabilized by a novel state labeled ``AF(2)'' \cite{kink}. 
The metallic vs. insulating character of the various phases can be 
analyzed with the Drude weight $\rm D_W$ \cite{previous}.
Results at $\lambda$=1.2 (Fig.~2(b)) 
show that the FM-state is metallic ($\rm D_W\ne$0) and the CE-state is 
insulating ($\rm D_W$=0). 
The jump in $\rm D_W$ again signals a first-order transition. 
Based on results similar to those in Figs.~2(a) and (b), the phase
diagram in 2D and low-T was constructed (Fig.~2(c)).
It presents a rich structure including FM and CE-phases.
In particular, it contains a CD FM-state with
{\it uniform} orbital order,
which after consideration of three dimensional (3D) 
clusters will become a fully A-type state with charge and orbital 
characteristics as observed experimentally \cite{PSMO}.
Then, all the currently known states at x=0.5 exist in our phase diagram, 
and they are in close competition.
It is conceivable that small temperature, doping, or lattice spacing changes 
may alter their balance, as found experimentally. 
It is also possible that the other novel phases in Fig.~2(c),
especially the CO and OO FM-state, may appear 
in future experiments under suitable conditions.

\begin{figure}[htbp]
\vspace{-0.5cm}
\centerline{\psfig{figure=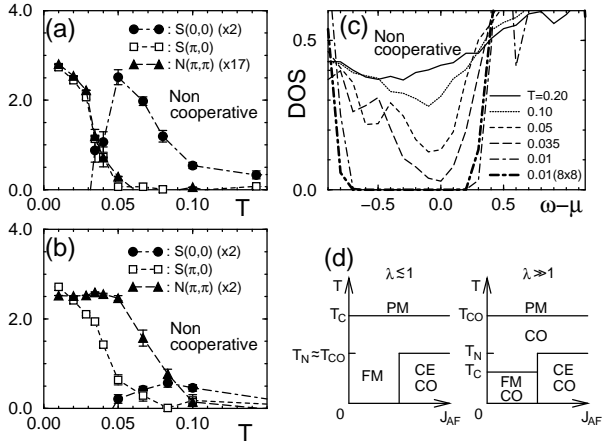,width=8.5cm,angle=-90}}
\vspace{0.4cm}
\caption{
(a) Spin structure factors S(0,0) and S($\pi$,0) and charge structure factor 
N($\pi$,$\pi$) vs. T for a $4\times4$ cluster with $\lambda$=1.2 and 
$\rm J_{AF}$=0.15. 
The high-T background, corresponding to on-site correlations only,
is subtracted. 
(b) Same as (a) but at $\lambda$=1.9 and $\rm J_{AF}$=0.10.
(c) DOS vs. $\omega$ for $4\times4$ and $8\times8$ clusters 
for $\rm J_{AF}$=0.15 and $\lambda$=1.2.
(d) Schematic phase diagram in the $\rm (J_{AF}, T)$ plane for 
the small $\lambda$ (left) and large $\lambda$ (right) regions.
PM means paramagnetic phase.
}
\vspace{-0.1cm}
\end{figure}

Here the cooperative effect on the above results is briefly discussed.
When the oxygen positions $\rm \{ u^v_{\bf i}\}$, not $\rm \{ Q \}$,
are used as variables in the MC simulations,
the charge, spin, and orbital correlations 
are found to be similar to the non-cooperative results.
In particular, a clear evidence of a CO CE-type state was observed 
in a wide range of the phase diagram and to stabilize this state
$\rm J_{AF}$ was found to play an essential role as in 
the non-cooperative case.
For $\rm J_{AF}$=0, a CO-state is still obtained 
at large $\lambda$, but it is spin FM.
In addition, note that the A-type phase of Fig.~2(c) has Q$_3$ modes 
with an {\it uniform} mean-value at every site. 
This amount to a static deformation of the oxygen octahedra, 
shortening the c-direction (or elongating the a- and 
b-directions), similarly as in experiments \cite{akimoto}.
If appropriate lattice constants are not used for cooperative phonons, 
the A-type phase may not be stabilized.
For this reason studying both non-cooperative and
cooperative phonons is important.

It is instructive to calculate the T-dependence of the various 
order parameters to compare with experiments. 
In Fig.~3(a), $\rm S({\bf q})$ at ${\bf q}$=(0,0) (FM) and 
($\pi$,0) (CE) and charge structure factor $\rm N(\pi,\pi)$ vs. 
T are shown for an intermediate $\lambda$.
At T$\sim$0.1t, S(0,0) grows with reducing T as if FM were to become stable. 
However, at $\rm T_{CO}$$\sim$$\rm T_{N}$$\sim$0.05t a $sudden$ transition 
to a CO CE-state is found (T$_{\rm CO}$$\sim$250K for t$\sim$0.5eV).
Note that the appearance of ferromagnetism in a temperature 
window above T$_{\rm CO}$ is in quantitative agreement
with results observed in $\rm La_{0.5} Ca_{0.5} Mn O_3$ (LCMO) \cite{schiffer},
and in $\rm (La_{1-z} Nd_{z})_{0.5} Sr_{0.5} Mn O_3$ (LNSMO) 
for z$\geq$0.6 \cite{akimoto}.
The results at large $\lambda$ are {\it qualitatively} different (Fig.~3(b)). 
Here T$_{\rm CO}>$T$_{\rm N}$, and only a weak FM signal is observed. 
These results are in good agreement with experiments for 
$\rm La_{0.5} Sr_{1.5} Mn O_4$\cite{sternlieb} and 
$\rm Pr_{0.5} Ca_{0.5} Mn O_3$\cite{tomioka}.
No metallic FM-state exists in these compounds, and our analysis predict 
that they have a larger $\lambda$ than LCMO and LNSMO \cite{comment1}.
The density of states (DOS) at several T's are in Fig.~3(c).
A robust charge gap $\Delta_{\rm CO}$$\sim$0.9t at 
T$\sim$0.01t is clear, as well as a precursor pseudogap at higher T's.
The ratio $\Delta_{\rm CO}$/$\rm k_B T_{CO}$$\sim$18 is remarkably close
to the tunneling results ($\sim$19) for NSMO~\cite{biswas}. 

Let us understand intuitively the several transition temperatures. 
In the small $\lambda$ region, the FM transition temperature $\rm T_C$
scales with t due to the double exchange mechanism, while 
$\rm T_N$ is determined by $\rm J_{AF}$ irrespective of $\lambda$.
In the FM phase, a CO-state never occurs in this region.
However, if a sharp transition to the CE-type AF state occurs at 
T=$\rm T_N$, 
the CO-state is also generated, since the CE-type state is 
a 1D band-insulator\cite{hotta} and 
it can be shown using a standard mean-field approximation that
$\rm k_{B}T_{CO}$$\approx$$\rm \varepsilon_{F}$$e^{-1/\lambda}$,
where $\rm \varepsilon_{F}$ is the Fermi energy 
much larger than $\rm J_{AF}$.
Thus, $\rm T_{CO} \sim T_N$ holds for small $\lambda$.
Note that the ratio $\Delta_{\rm CO}/\rm k_B T_{CO}$ is large 
because $\Delta_{\rm CO}$$\sim$t in the band-insulating region.
When $\lambda$ is increased, 
$\Delta_{\rm CO}$ and $\rm T_{CO}$ smoothly increase,
keeping the relation $\Delta_{\rm CO}$$\gg$$\rm k_{B}T_{CO}$.
Eventually, at large $\lambda$ $\rm T_{CO}$ overcomes $\rm T_N$.
In this regime, $\rm T_C$ is determined by the superexchange 
mechanism as $\rm T_C$$\sim$$\rm t/\lambda^2$.
The above estimations are summarized in Fig.~3(d).
In the small $\lambda$ region,
the metallic-like FM phase appears irrespective of $\rm J_{AF}$
at high T, but at low T, the CE-type AF state
occurs for intermediate value of $\rm J_{AF}$
and the CO-state is concomitant only to the CE-phase.
In the large $\lambda$ regime, the CO-state occurs in the whole region 
for T$<$$\rm T_{CO}$, higher than $\rm T_C$ and $\rm T_N$.
At low T, the system becomes FM-insulator or 
CE-type insulator depending on $\rm J_{AF}$.
All these results are supported by the MC simulations.
Note that previously it was widely believed that 
modifications in the tolerance factor $\rm t_F$ by chemical substitution 
lead to bandwidth changes
sufficiently strong to drastically affect the properties of
manganite compounds. However, our results established that
the $\rm t_F$-induced changes in $\rm J_{AF}$ are equally
important, and they should be carefully considered when studying Mn-oxides.

A difficult to understand feature of the manganite 
CO-state is the $z$-axis charge stacking. 
To address the prediction of JT-phonon studies, 
in Fig.~4(a) the MC energy of a bilayer cluster vs. $\rm J_{AF}$ 
with non-cooperative phonons is shown. 
At small $\rm J_{AF}$ the FM-state dominates, as in 2D (Fig.2(c)).
The CE-state in its ``Wigner Crystal'' (WC) and ``Charge Stacked'' (CS)
versions (with a $\rm N({\bf q})$ peak at $(\pi,\pi,\pi)$ and
$(\pi,\pi,0)$, respectively) appear
as excited states at $\rm J_{AF}$=0.
WC has lower energy since the electronic kinetic energy
improves when the Mn$^{3+}$-ions spread apart.
However, with increasing $\rm J_{AF}$ the
situation rapidly changes since the CS charge arrangement
(CE-planes, one over the other, with opposite spins) has all $z$-axis bonds 
properly antiferromagnetically aligned.
On the other hand, in the WC arrangement, with one plane shifted from
the other by a lattice spacing along x or y, half the $z$-axis bonds 
are AF and the other half FM, making its energy $\rm J_{AF}$-independent. 
For this reason, at realistic $\rm J_{AF}$, the CS-state dominates.
Since similar results are obtained 
with cooperative phonons and cubic clusters (Fig.4(b)),
the effect appears robust and independent of fine details. 
The important interaction to stabilize the CS-state is the $z$-axis 
$\rm J_{AF}$ exchange.

\begin{figure}[htbp]
\vspace{-0.60cm}
\centerline{\psfig{figure=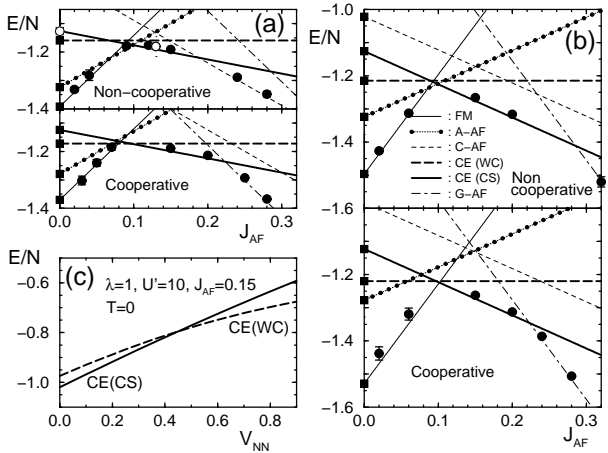,width=8.5cm,angle=-90}}
\vspace{0.5cm}
\caption{
(a) Energy per site vs. $\rm J_{AF}$ at $\lambda$=1.5, $\rm J_H$=$\infty$,
and T=1/100 with both cooperative and non-cooperative phonons.
Solid (open) circles are MC results using a 4$\times$4$\times$2 
(8$\times$8$\times$2) bilayer. 
Squares at $\rm J_{AF}$=0 are the MC results on the 4$\times$4$\times$2 
cluster assuming the $\rm t_{2g}$-spin background according to the convention 
in (b), in a standard notation.  
The straight lines were obtained as in Fig.~2(a).
(b) Same as (a) but for a cube 4$^3$. 
Note that with non-cooperative phonons the A-type phase is stable in 
bilayers and cubes in a narrow $\rm J_{AF}$-window.
(c) Mean-field energy per site vs. $\rm V_{NN}$ for a bilayer. 
Couplings are realistic ($\rm U'$ is the 
on-site inter-orbital repulsion).
Since $\rm J_H$ is assumed infinite,
the on-site intra-orbital repulsion is irrelevant.
Solid and dashed curves denote the CE-type CS- and 
and WC-structures, respectively. 
}
\vspace{-0.1cm}
\end{figure}

What is the influence of the $\rm V_{NN}$ repulsion in our results?
It will certainly penalize the CS structure, and the WC-state will 
eventually become the ground state as this repulsion grows.
However, the CS-state has a better magnetic energy than the WC-state.
Comparing the energy per site gained (-$\rm J_{AF}$) and lost ($\rm V_{NN}/2$)
in the 3D CS structure, a critical coupling 
$\rm V^c_{NN}$=$\rm 2J_{AF}$$\sim$$\rm 0.15$eV is estimated 
if $\rm J_{AF}$$\sim$0.15t and t$\sim$0.5eV.
A more sophisticated  mean-field approximation\cite{mean-field}
at realistic manganite couplings provides 
$\rm V^c_{NN}$$\sim$$\rm 0.22$eV (Fig.~4(c)).
Although the $\rm V_{NN}$ bare value is $\sim$3.6eV,
it is reduced by the dielectric constant $\epsilon \approx 20$-$45$
\cite{arima}, leading to $\rm V_{NN}$$\approx$0.08-0.18eV for manganites.
Thus, it is concluded that the manganite 
NN repulsion is weak enough to allow for the CS-state to be
stable due to the effect of $\rm J_{AF}$.

Summarizing, evidence was provided that the CE-type CO-state with 
$\Delta_{\rm CO}$/${\rm k_B T_{\rm CO}}$$\gg$1 and charge stacked along the 
$z$-axis observed at x=0.5 in several manganites can be stabilized 
using JT phonons.
The competing FM and A-type states also appear in the simulations.
Our results have established that purely Coulombic approaches are not the 
only procedure to stabilize CO-states in manganites, but strongly coupled
electron JT-phonon systems provide an alternative framework where controlled
many-body calculations are possible.

TH is supported from the Ministry of Education, Science, Sports, and Culture 
of Japan. ED is supported in part by grant NSF-DMR-9814350.



\begin{references}
\bibitem{comm6} Only at large temperatures tunneling between the two 
possible CO-states is observed in the MC analysis.
\bibitem{kink} The AF(2)-state has $\rm t_{2g}$-spins arranged as 
$\rm \uparrow \uparrow \downarrow \downarrow$ periodically along the x-axis. 
Along the y-axis the spins are AF.
\bibitem{mean-field} In our formalism, to be described elsewhere,
the electron-phonon and electron-electron terms are Hartree-Fock decoupled, 
but the orthogonal bases for $e_g$ electrons at each site is self-consistently
determined.
\end{references}
\end{document}